\documentclass[12pt]{article}
\usepackage{graphics}  % import .ps files
\usepackage{amsmath}   % math symbols
\usepackage{times}     % Times-Roman font
\usepackage{authblk}   % author affiliation

\newcommand{\nanohub}{nanoHUB}
\newcommand{\polymod}{PolymerModeler}
\newcommand{\degrees}{$^{\circ}$}

\begin{document}

%\begin{frontmatter}

%\title{Modeling amorphous polymers at \nanohub.org}
\title{Atomistic simulations of amorphous polymers in the cloud with \polymod}

\author[1]{Benjamin P. Haley}
\author[2]{Chunyu Li}
\author[2]{Nathaniel Wilson}
\author[3]{Eugenio Jaramillo}
%\author[2]{Alejandro Strachan\corref{author}}
\author[2]{Alejandro Strachan\thanks{strachan@purdue.edu}}
\affil[1]{Research Computing}
%\cortext[author] {Corresponding author. \textit{E-mail address:} strachan@purdue.edu}
\affil[2]{School of Materials Engineering and Birck Nanotechnology Center\\Purdue University, West Lafayette, Indiana 47907}
\affil[3]{Department of Biology and Chemistry\\Texas A\&M International University, Laredo, TX 78041}

\maketitle

\begin{abstract}
Molecular dynamics (MD) simulations enable the description of material properties and processes with atomistic detail by numerically solving 
the time evolution of every atom in the system. We introduce \polymod, a general-purpose, online simulation tool to build atomistic 
structures of amorphous polymers and perform MD simulations on the resulting configurations to predict their thermo-mechanical properties. 
\polymod\ is available, free of charge, at \nanohub.org via an interactive web interface, and the actual simulations are performed in the cloud 
using  \nanohub.org resources. Starting from the specification of one or more monomers \polymod\ builds the polymer chains into a 
simulation cell with periodic boundary conditions at the desired density. Monomers are added sequentially using the continuous configuration 
bias direct Monte Carlo method, and copolymers can be created describing the desired sequence of monomers. \polymod\ also enables users 
to perform MD simulations on the structures created by the builder using the Dreiding force field and Gasteiger partial atomic charges.  We 
describe the force field implementation and the various options for the MD simulations that use the LAMMPS simulator. 
\polymod\ MD simulations for a PMMA sample show good structural agreement with experiments and are in good 
agreement with simulations obtained with commercial software.
\end{abstract}

%\begin{keyword}
%amorphous polymer \sep configurational bias methods \sep molecular dynamics \sep \nanohub\ \sep online simulations
%\end{keyword}

%\end{frontmatter}

%\linenumbers

% -----------------------------------------------------------------------------
\section{Introduction}

Atomistic simulations are a powerful tool to predict materials properties and uncover the fundamental mechanisms that govern them from first
principles; these techniques are expected to play an increasingly important role in the design of new materials with tailored properties. 
\cite{ref_MGI,Strachan:2005p29}  Among many other contributions, molecular dynamics (MD) simulations have provided invaluable insight into the 
response of materials under extreme conditions\cite{Kadau:2002p4168,Holian:1998p538}, size effects on plastic deform\-ation\cite{Schiotz:2003p5411}, 
and polymer dynamics \cite{Warren:2010p6645}. MD simulations are ubiquitous in science and engineering due, in part, to the continuing increase in 
computing power and the availability of efficient, parallel, and scalable codes that  can make effective use of current supercomputers 
\cite{Plimpton:1995p795,Phillips:2005p7881,Smith:2005p7885,Brown:2011p7702,ref_MOLDY}.

Even though some of these simulation codes, such as ESPResSo\cite{ref_ESPRESSO}, 
\\GROMACS\cite{ref_GROMACS}, and AMBER\cite{ref_AMBER} are designed to simulate soft matter systems, the simulation of amorphous polymers remains challenging due to the difficulties involved in generating accurate molecular structures to be used as initial conditions in MD simulations.  Unlike crystalline materials where initial atomic positions can be easily generated by replicating a unit cell, or in {\it atomic} amorphous solids where equilibrated structures can be obtained from the melt \cite{Anderson:2011p7415,VedulaPRB}, the generation of amorphous thermoset or thermoplastic polymer structures requires special-purpose algorithms \cite{Sadanobu:1997p9805,Theodorou:2004p6109,Li:2011p10031,Varshney:2008p1095,Li:2010p7210,Colina2013,Colina2014}. Chain dynamics in linear-chain polymers involve timescales well beyond those achievable with MD simulations, especially for large molecular-weight cases\cite{ref_POLYMER_TEXT}, and if the initial structure does not exhibit correct statistical properties the simulations will produce inaccurate results.
\par
Some commercial software packages, such as MAPS by Scienomics and Materials Studio by Accelrys, Inc, offer the capability to build atomistic structures of condensed-phase amorphous polymers using a simple GUI, pack the chains into a simulation volume, and perform MD simulations. Open source codes designed to build amorphous polymers include Polymatic \cite{Colina2013,Colina2014}, 
MCCCS Towhee\cite{ref_MCCCS,ref_MCCCS_web}, PACKMOL\cite{ref_PACKMOL}, AMBER\cite{ref_AMBER}, and GROMACS\cite{ref_GROMACS}.  
These codes requires users to create input files in domain-specific mini-languages, in addition to compiling and running the program in 
order to pack chains in preparation for MD simulations.  The ECCE\cite{ref_ECCE} tool offers a GUI for building 
systems that may be simulated using \textit{ab initio} packages such as Gaussian\cite{ref_GAUSSIAN} but not with 
general purpose MD codes.
Thus, we designed and developed \polymod\  to provide a easy-to-use, general-purpose, freely accessible tool to build amorphous thermoplastic structures and perform MD simulations. We expect the tool to be useful both for education and in research: setting simulations up is simple enough to be used in the classroom, all that is required is an internet connection, yet, the code is powerful enough for research and can also be used in conjunction with state-of-the-art techniques to relax polymer structures and predict their properties.
\par
The \polymod\ tool has been deployed on \nanohub.org and is freely available for online simulations via the URL http://nanohub.org/resources/polymod.  The only requirement is a \nanohub\ account that can be obtained after a free registration.  \nanohub\ \cite{ref_NANOHUB} makes simulation tools accessible through a standard web browser or an iPad app, removing the need for users to download, configure, and install software and learn scripts or domain-specific languages.  The state of a simulation is preserved even after closing the browser in which it is running or logging out of \nanohub\ so that the user can return to the simulation later. Users can provide feedback and ask questions in \nanohub.org.  This information has guided the development of \polymod, and we will continue listening to the desires of the user community for future development.
\par
The \polymod\ tool features a builder of amorphous, linear-chain thermoplastic polymers, detailed in Section \ref{sec_builder}.  As described in Section \ref{sec_dynamics}, the atomistic system constructed by the builder can be used as initial conditions for MD simulations using LAMMPS\cite{ref_LAMMPS,ref_LAMMPS_WEB} to compute thermo-mechanical properties of the polymer. §All simulations run on \nanohub\ computing resources including the possibility of running the MD in parallel over multiple processors. Conclusions are drawn in Section \ref{sec_conclusions}.

% -----------------------------------------------------------------------------
\section{Amorphous polymer builder}
\label{sec_builder}

The process of building a condensed-phase, amorphous, linear-chain polymer in \polymod\ can be divided in three main steps: i) the specification of one or multiple monomers and their statistical arrangement into chains, including the chain length and number of chains to pack in the simulation cell, ii) the determination of the simulation cell into which the chains will be placed; and iii) the specification of rules for the sequential placement of monomers into chains of the desired length with conformation given by a set of dihedral angles between monomers. In \polymod\ this last step is performed using the continuous configurational bias Monte Carlo algorithm \cite{Sadanobu:1997p9805}. The following subsections describe these steps including details of the algorithms, approximations, and their implementation after which results of the builder are presented. Once the chains are built, \polymod\ performs a series of relaxations to remove bad contacts and reduce entanglements after which molecular dynamics can be performed to predict thermo-mechanical properties; these relaxation and MD steps are described in Section \ref{sec_dynamics}.

\subsection{Monomers and their arrangement}
\label{sec_monomer}
The first step in building amorphous thermoplastic polymers is specifying the repeating unit which will be used to construct chains, and users can choose between a set of pre-built monomers or create/upload their own.  The \polymod\ tool accepts monomer specifications in the commonly used PDB (protein data bank) and XYZ file formats. Files of these types are available from a wide variety of online databases and may be uploaded directly to the \polymod\ tool with no modification.  The user must also specify which atom in the monomer is the head atom and which is the tail.  The builder constructs a chain by adding one monomer unit at a time.  Once a new monomer is added, the tail atom of the previous monomer on the chain is removed and so is the head atom of the new monomer.  The final step is the creation of a bond between the backbone atoms originally attached to the head and tail atoms just removed. The head and tail atoms must each be connected to the molecule by a single bond.
\par
Internally the \polymod\ tool converts a PDB or XYZ monomer to a z-matrix (internal coordinates format) representation that details the bonds, bond lengths, bond angles, and torsion (dihedral) angles between atoms.  In this representation, the head atom is listed first, followed by all the backbone atoms to the tail, then all other atoms in the monomer. The \polymod\ tool also accepts monomer specifications in this z-matrix format. The box describes the z-matrix format and shows an example for a PMMA monomer.  

% ---------------------------------------------------------
\begin{figure}[h]\begin{center}
\scalebox{0.7}[0.7]{\includegraphics{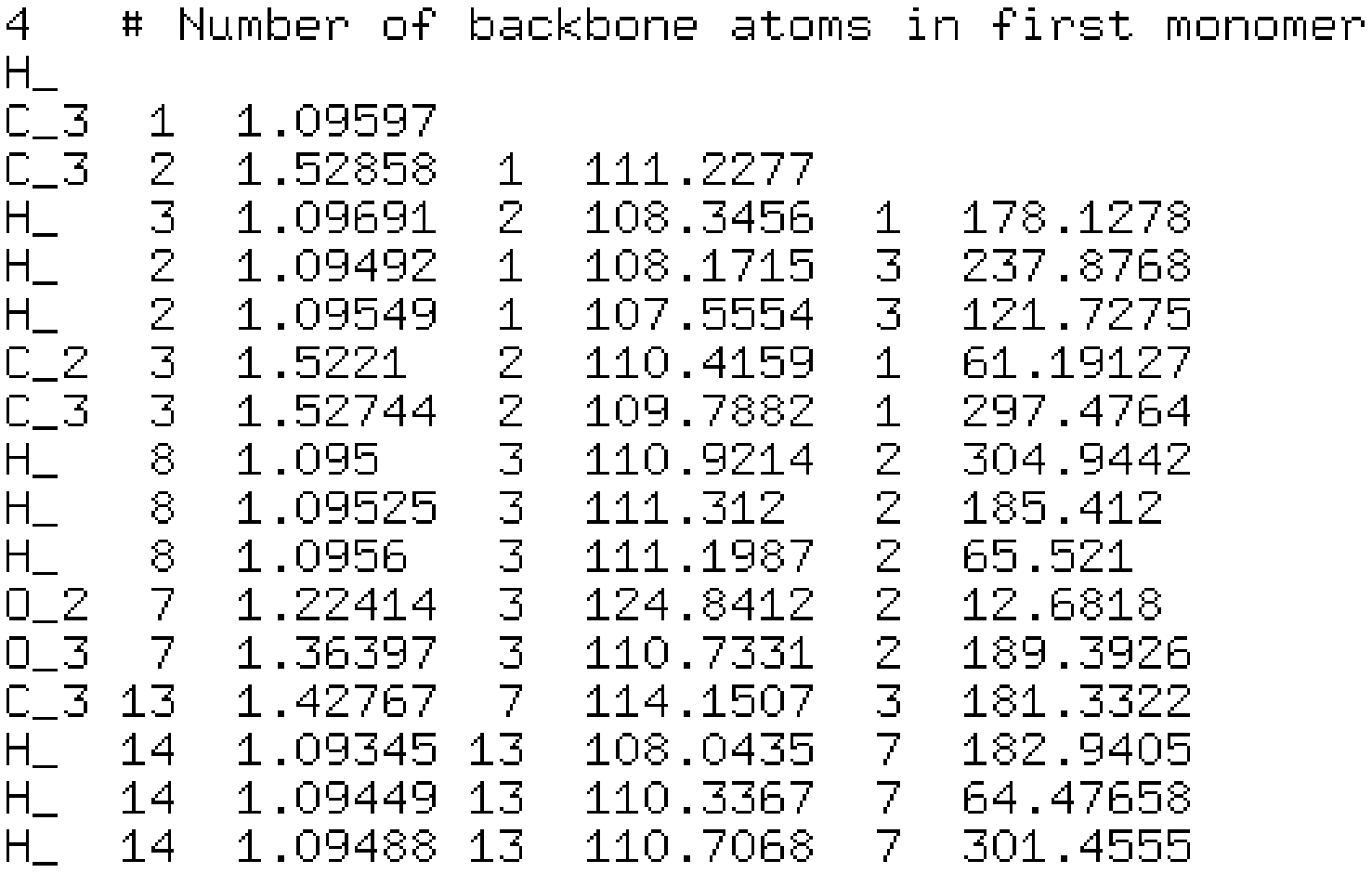}}
\caption{\label{fig_zmatrix}Internal coordinate representation of a PMMA 
repeated unit}
\end{center}
\end{figure}

\fbox{\begin{minipage}{12cm}
{\bf The z-matrix format}

The first line specifies the number of backbone atoms in the monomer, which are the leading entries in the z-matrix.  The first column of the z-matrix gives information about atomic species in terms of Dreiding atomic types (discussed in the following paragraph). 
From the second atom on, the second column of the z-matrix indicates the index of a previously defined atom to which the current atom (the atom specified by the current line) is bonded; the third column is then the bond length, in Angstroms, between these two atoms.   For the third atom on, an angle needs to be specified to fully determine its internal position, so two additional columns are needed indicating a third atom (column four) and the corresponding bond angle, in degrees, is in the fifth column.  For the remaining atoms a dihedral angle needs to be specified; the sixth column gives the index to a fourth atom required to specify a dihedral angle (defined between the plane determined by atoms on columns 6, 4 and 2 with that determined by atoms on columns 1, 2 and 4).  The torsion angle between these planes is given, in degrees, in the final column.  Pre-built z-matrix representations of several monomers are available in \polymod.  

\end{minipage}}

Once the list of monomers are specified, the \polymod\ builder accepts a list of monomer arrangements, consisting of repeating patterns or probabilities to create random structures. A pattern arrangement specifies an order in which monomers are added to a chain, repeating until the desired chain length is reached.  A probability arrangement assigns relative weights to the defined monomers, which are stochastically added to a chain according to these weights.  Multiple arrangements (patterns or probabilities) can be defined, with relative weights, to support the growth of multiple chain configurations in the same system.  The builder chooses a pattern for each chain added to the system, according the the weights assigned to the patterns, and constructs the entire chain according to the selected pattern.  In this manner the builder allows the full specification of stereochemistry. The supplementary material provides examples of how these features can be used to create isotactic, syndiotactic, and atactic PMMA structures. 

\subsection{Simulation cell, molecular weight and number of chains}
%\label{}
The remaining parameters that define the amorphous cell are the number of chains, built from to the monomer arrangements described in section \ref{sec_monomer} (with configurations determined by methods detailed in section \ref{sec_conformations}), the number of monomers in each chain (that determines molecular weight), the size of the simulation cell, and the temperature.  The simulation cell size can be specified directly, as the lengths of a orthogonal cell along the three Cartesian directions (in Angstroms) or as the density of a cubic box, whose dimensions are determined from the total mass and the density.

\subsection{Configuration building}
\label{sec_conformations}

The principal polymer builder algorithm supported by \polymod\ is the continuous configurational biased direct Monte Carlo\cite{Sadanobu:1997p9805} that efficiently samples torsions along the polymer backbone and leads to structures with reasonable statistical weights. Chain conformations are governed by covalent interactions associated with torsional angles along the backbone and by non-bond interactions that cause chains to avoid overlaps and for affine groups to tend to coalesce. These interactions cause chains to extend in good solvents and coil in bad solvents. In addition to the continuous configurational biased direct Monte Carlo algorithm \polymod\ allows to build freely rotating chain polymers, rod-like polymers and chains built using self-avoiding walks for educational and verification purposes. 

\subsubsection{Continuous Configurational Biased Direct Monte Carlo}

Due to their high stiffness, bond lengths and angles are fixed at their equilibrium values throughout the building procedure bond distances; this is commonly done in polymer builders. Thus, dihedral angles between backbone atoms ($\phi_i$) are the only degrees of freedom considered during the initial structure creation. The first monomer is placed into the simulation cell at a random location (with uniform probability density) and with random orientation (with uniform angular distribution). Monomers are added sequentially with their torsion angles chosen according to the Boltzmann weight, i.e. the probability density of torsion $i$ taking the angle $\phi_i$ is proportional to $\exp\left[-E\left(\phi_i\right)/kT\right]$ where $E\left(\phi_i\right)$ is the energy of the of the corresponding configuration, $k$ is Boltzmann's constant and $T$ is the absolute temperature. The energy associated with the configuration is obtained by adding van der Waals interactions and the covalent torsional energy as described in the following sub-sections.

The \polymod\ builder uses a Metropolis\cite{ref_METROPOLIS} Monte Carlo algorithm to explore and select torsion angle configurations.  Initial angles are assigned randomly for all flexible torsions in the new monomer and a Metropolis Monte Carlo Markov chain is performed to determine the actual torsion angle values according to their Boltzmann weight. The Monte Carlo steps are as follows:

\begin{itemize}
\item A new trial configuration (torsion angles for all flexible dihedrals of the current monomer) is chosen by an unbiased, random modification of the current state, 
\item The change in energy, $\Delta E$, associated with the trial move is computed from the van der Waals interactions and covalent torsion terms (if specified);
\item The new configuration is accepted with probability one if $\Delta E<0$ (i.e. the trial move decreases the energy of the system) and with probability $e^{-\frac{\Delta E}{k_BT}}$ otherwise.
\end{itemize} 
This process is repeated for a user-specified number of steps, and the final angles are used to place the monomer.
\par

{\it Covalent energy of backbone torsion angles} The \polymod\ amorphous builder allows users to specify the covalent torsional potential in two ways. The torsional energy as a function of angle can be provided numerically. Sample torsion potentials for sp$^2$ and sp$^3$ bonds according to the Dreiding force field are included in the tool. Alternatively, users can specify a probability as a function of torsion angle; these probabilities are then converted into relative energies using Boltzmann weights. Energies for arbitrary atoms are obtained by interpolating between the tabular data.

{\it Non-bond interactions.} Non-bond interactions in the continuous configurational biased direct Monte Carlo play a key role avoiding bad contacts between atoms. \polymod\ computes the non-bond energy as the sum of Lennard-Jones pair interactions between each atom in the tail monomer and all surrounding atoms, using Dreiding parameters. As is customary in molecular force fields, atoms in the same chain separated by fewer than four bonds are excluded from these non-bond interactions.  The cutoff number of bonds for non-bond interactions is a configurable parameter.

{\it Computation of pair-wise interactions}. A brute-force calculation of interatomic distance between the atoms of every new monomer with all previously built requires $1/2N(N-1)$ computations (N being the number of atoms) and becomes prohibitively expensive for relatively large systems. Thus, \polymod\ uses the well-known domain decomposition approach to transform this computation into an $O(N)$ calculation. The simulation cell volume is divided via a 3D grid with spacing $\Delta L$, close to but larger than the cutoff distance for van der Waals interactions for the Monte Carlo algorithm. As new monomers are added into the system atoms are placed into their corresponding cell; this involves $O(N)$ calculations. In order to compute inter-atomic distances to compute van der Waals energies or impose volume exclusions for an atom belonging to cell $i$ only atoms in cell $i$ and in its 26 nearest neighbors need to be checked. This calculations is independent of the simulation cell size for a given density and, thus, the algorithm is $O(N)$.

\subsection{Other Chain Builder Methods}

In addition to the configurational biased direct Monte Carlo, \polymod\ also allows users to build polymer according to the methods described in the following paragraphs which are included for educations and verification purposes alone.

{\it Freely rotating chains.} In this case all torsional angles have equal probabilities regardless of the covalent or van der Waals energies involved.

{\it Rod-like polymer chains} can be created by specifying a single torsional angle to be imposed on all torsional degrees of freedom.

{\it Self-avoiding chains with a hard cutoff} can also be created using excluded volumes. Excluded volumes are enforced by rejecting any
configuration that places any 2 atoms within a cutoff distance.
The attrition problems introduced by this approach are well known \cite{ref_CONFIG_BIAS,ref_INV_RESTRICT,ref_SMITH_FLEMING,ref_PERM} and we support it for educational and verification purposes.

\subsection{Amorphous builder results}
\label{sec_builder_results}
This section demonstrates various capabilities of the \polymod\ amorphous builder including output functions and verifies its implementation by comparing simulation results with analytical solutions for known cases and well-established solutions. 
%All results were obtained using the \polymod\ tool at \nanohub.org.

\subsubsection{Freely rotating chains}
\label{sec_freely_rotating}
\begin{figure}[h!]\begin{center}
\scalebox{0.5}[0.5]{\includegraphics{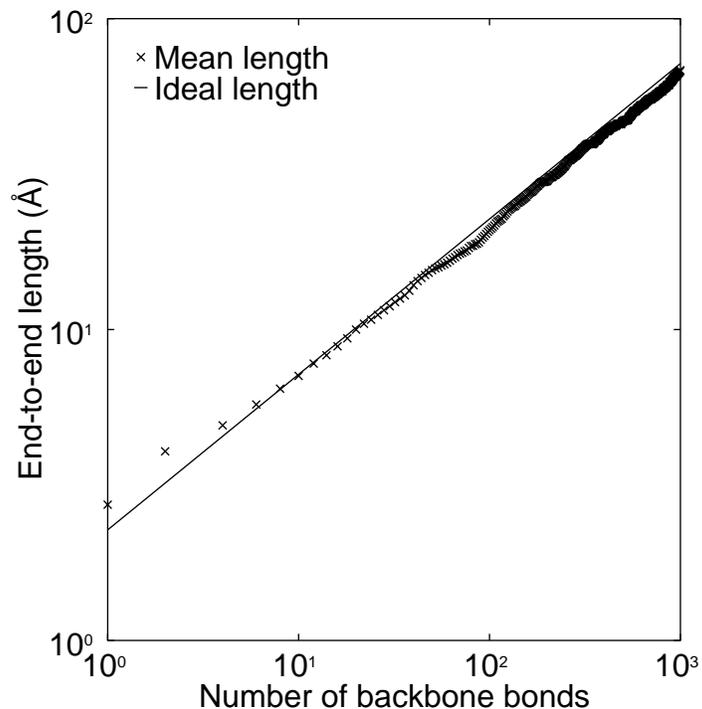}}
\caption{\label{fig_fr_mean_length}End-to-end length of 200 chains of freely 
rotating isotactic PMMA with 500 monomers each}
\end{center}
\end{figure}

\begin{figure}[h!]\begin{center}
\scalebox{0.5}[0.5]{\includegraphics{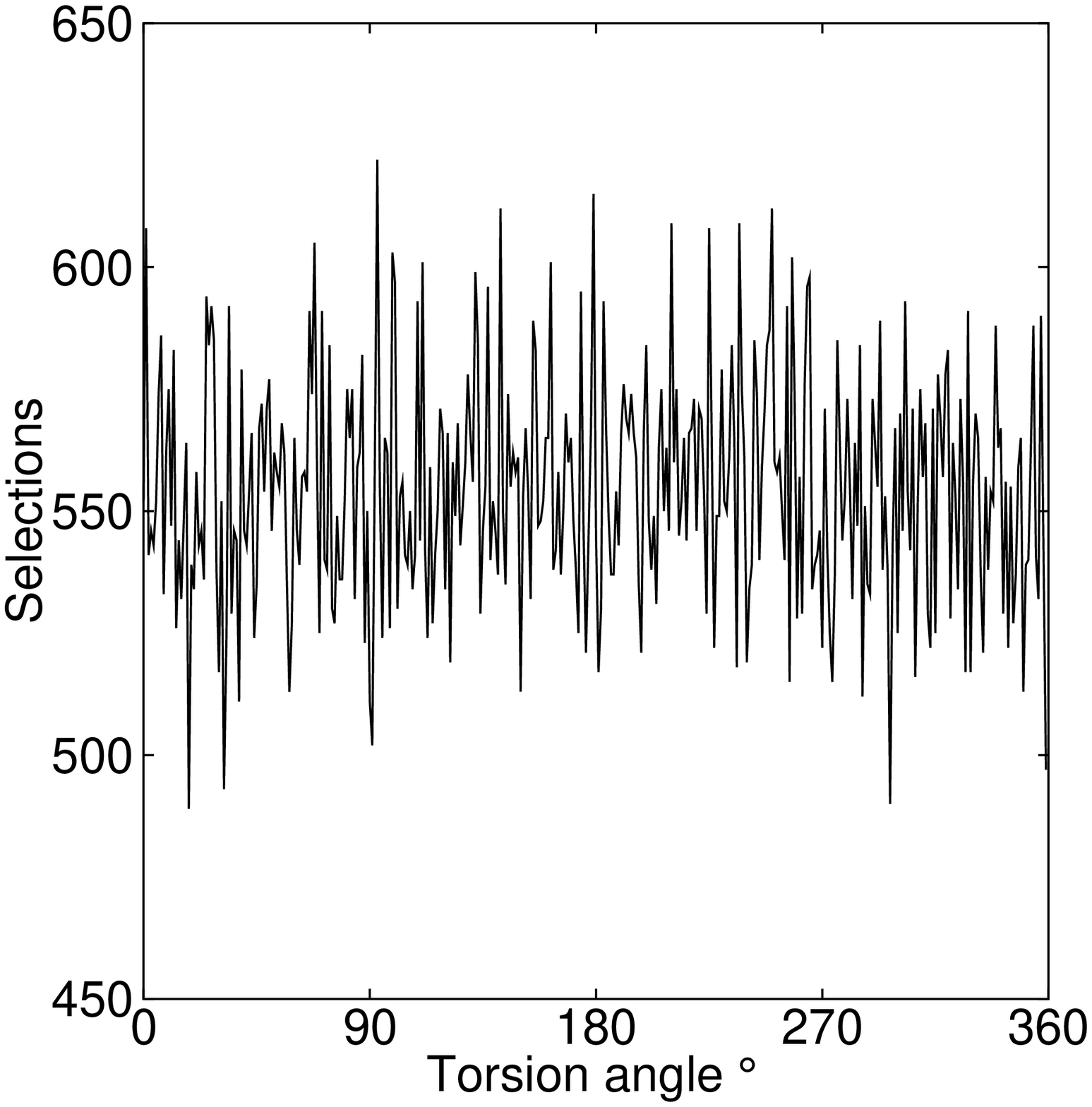}}
\caption{\label{fig_fr_torsions}Torsion angle histogram for 200 chains of freely
rotating isotactic PMMA with 500 monomers each}
\end{center}
\end{figure}

Figure \ref{fig_fr_mean_length} shows the mean end-to-end length as a function of number of backbone bonds $n$ obtained for 200 freely rotating chains of isotactic PMMA with 500 monomers in each chain. Each chain was built independently of the others, with only one chain in the simulation cell at any given time.  The \polymod\ result is compared with the exact solution that in this case is known analytically as it corresponds to a simple random walk \cite{ref_POLYMER_TEXT}:
% ---------------------------------------------------------
\begin{equation}
\label{eq_ideal_length}
\left<R\right> = l\sqrt{n\frac{1+cos\theta}{1-cos\theta}}
\end{equation}
% ---------------------------------------------------------
% ---------------------------------------------------------
with a C-C backbone bond length $l$ = 1.53{\AA} and backbone bond angle $\theta = 68$\degrees. The simulation result agrees with the analytical solution. Figure \ref{fig_fr_torsions} shows the distribution of torsion angle values for the same simulation. Torsions can take any value with equal probability in a freely rotating system and the code produces the expected result.
% ---------------------------------------------------------
% ---------------------------------------------------------
\par
\begin{figure}[h]\begin{center}
\scalebox{0.5}[0.5]{\includegraphics{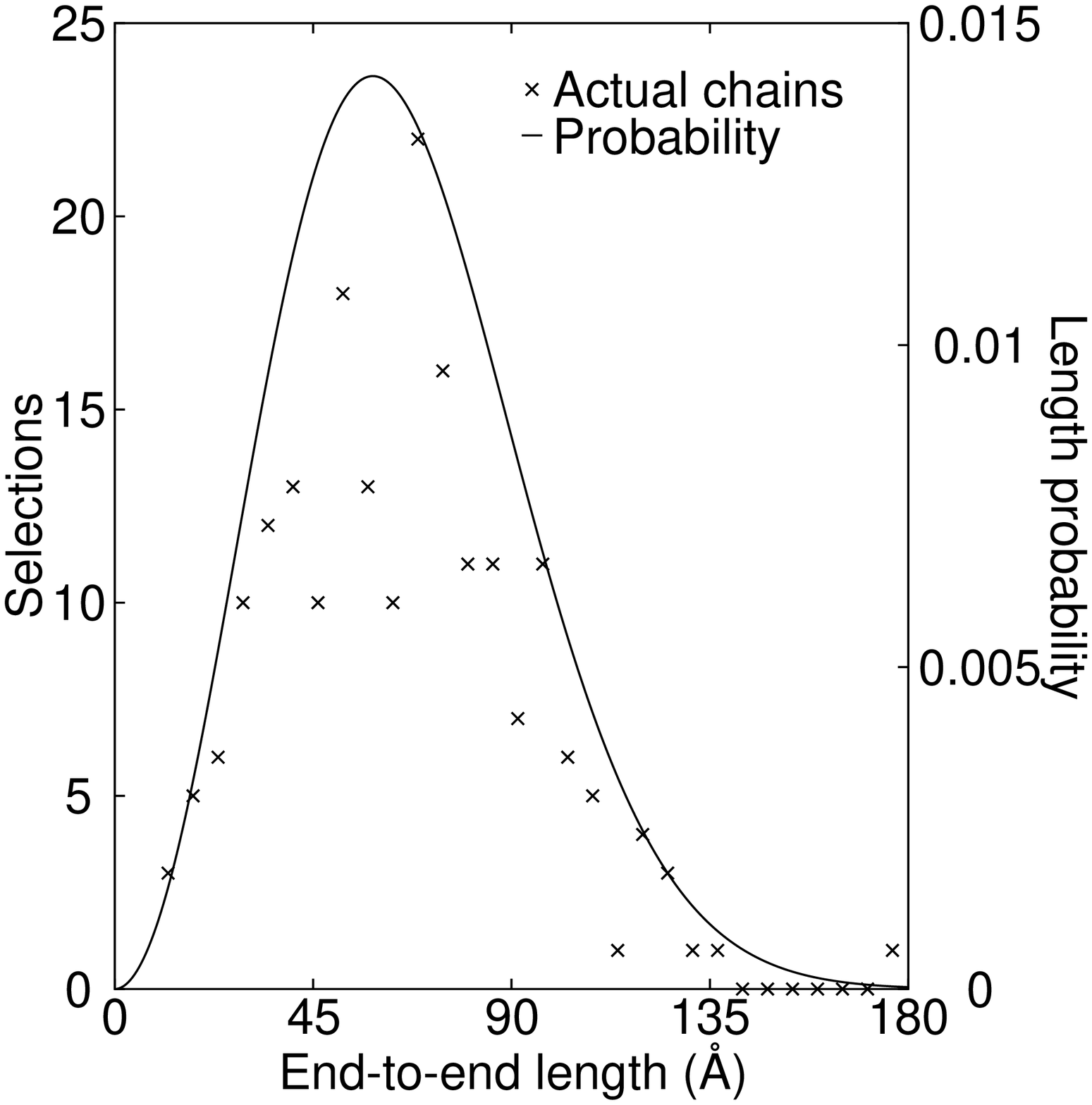}}
\caption{\label{fig_fr_length_histo}End-to-end length histogram for 200 chains 
of freely rotating isotactic PMMA with 500 monomers each}
\end{center}
\end{figure}

The distribution of end-to-end chain lengths for the same PMMA system is shown in Figure \ref{fig_fr_length_histo}.  The expected distribution \cite{ref_POLYMER_TEXT}, the probability of finding the end of a chain with length $\left<R\right>$ in a spherical shell with volume $4\pi R^2dR$, is also shown:
% ---------------------------------------------------------
\begin{equation}
\label{eq_length_prob}
P(R, \left<R\right>)4\pi R^2dR = 4\pi\left(\frac{3}{2\pi\left<R^2\right>}\right)^{\frac{3}{2}}\exp\left(\frac{-3R^2}{2\left<R^2\right>}\right)R^2dR
\end{equation}
% ---------------------------------------------------------
with $\left<R\right>$ calculated using (\ref{eq_ideal_length}). The simulation and theoretical result agree well, and using a larger sample of chains would
improve the agreement.
% ---------------------------------------------------------
% ---------------------------------------------------------

\subsubsection{Chains obtained via Continuous Configurational Biased Direct Monte Carlo}

In this section we present the results of building the same system as in the previous section, 200 chains of isotactic PMMA with 500 monomers in each chain, but using the continuous configurational bias Monte Carlo approach considering Lennard-Jones interactions between the atoms in the chains (no covalent interactions in this example). The cutoff range for the interactions is 4 {\AA}. Each chain is built independently, as in the freely rotating case, to facilitate comparisons with analytical results.   The Metropolis Monte Carlo sampling described in section \ref{sec_conformations} select torsions considering van der Waals interactions which avoids bad contacts, producing an effectively self-avoiding chain with a soft core.
% ---------------------------------------------------------
\begin{figure}[h]\begin{center}
\scalebox{0.5}[0.5]{\includegraphics{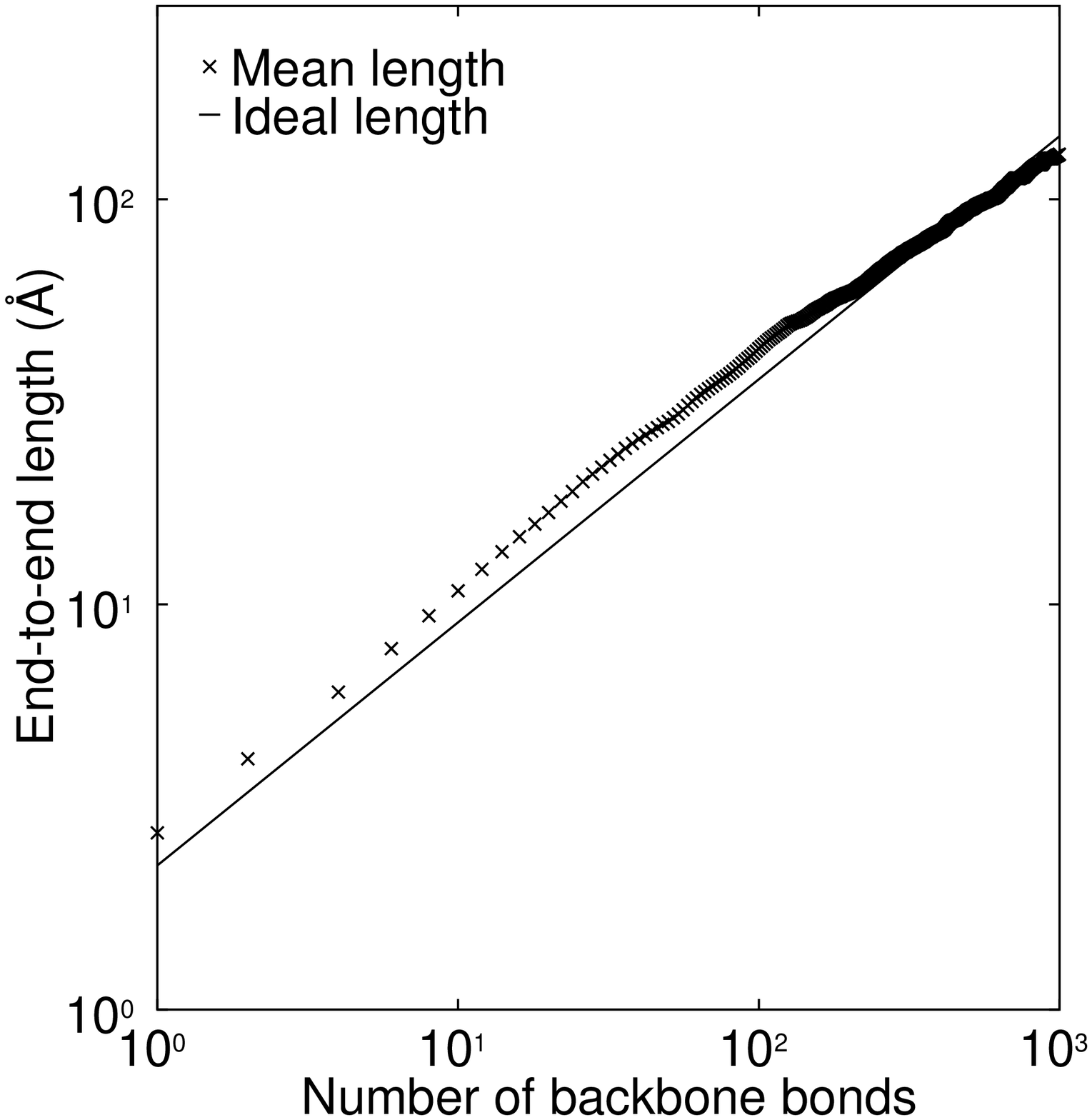}}
\caption{\label{fig_sa_mean_length}End-to-end length of 200 chains of self-avoiding isotactic PMMA with 500 monomers each}
\end{center}
\end{figure}
% ---------------------------------------------------------
\par
Figure \ref{fig_sa_mean_length} shows the end-to-end chain length of the chains in this system as a function of the number of backbone bonds, $n$. The theoretical trend\cite{ref_POLYMER_TEXT} in end-to-end length as a function of $n$ is similar to the random walk case (\ref{eq_ideal_length}) but increases faster than the square root of the number of monomers.  Self avoid walks show a $n^{0.6}$ dependency that is also shown in the figure, demonstrating good agreement between \polymod\ and theory.  As expected, the short-range repulsive interactions cause the chains to swell with respect to the freely rotating example, leading to longer chains.  
\par
% ---------------------------------------------------------
\begin{figure}[h]\begin{center}
\scalebox{0.5}[0.5]{\includegraphics{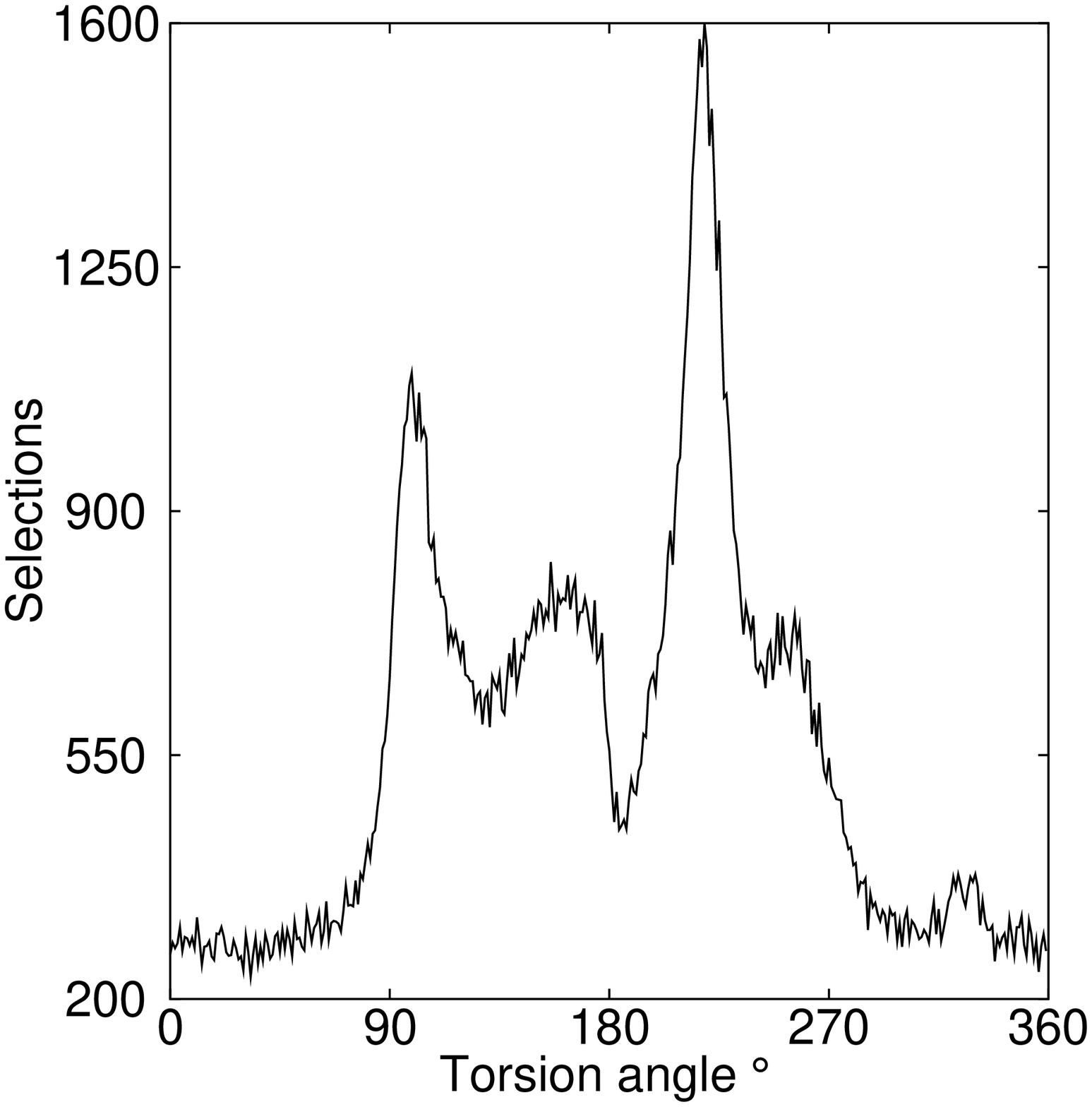}}
\caption{\label{fig_sa_torsions}Torsion angle histogram for 200 chains of 
self-avoiding isotactic PMMA with 500 monomers each}
\end{center}
\end{figure}
% ---------------------------------------------------------
The interactions also change the distribution of torsion angles, shown in Figure \ref{fig_sa_torsions}. Torsion angles between 90 and 270 degrees are more energetically favorable. 
\par
\begin{figure}[h]\begin{center}
\scalebox{0.5}[0.5]{\includegraphics{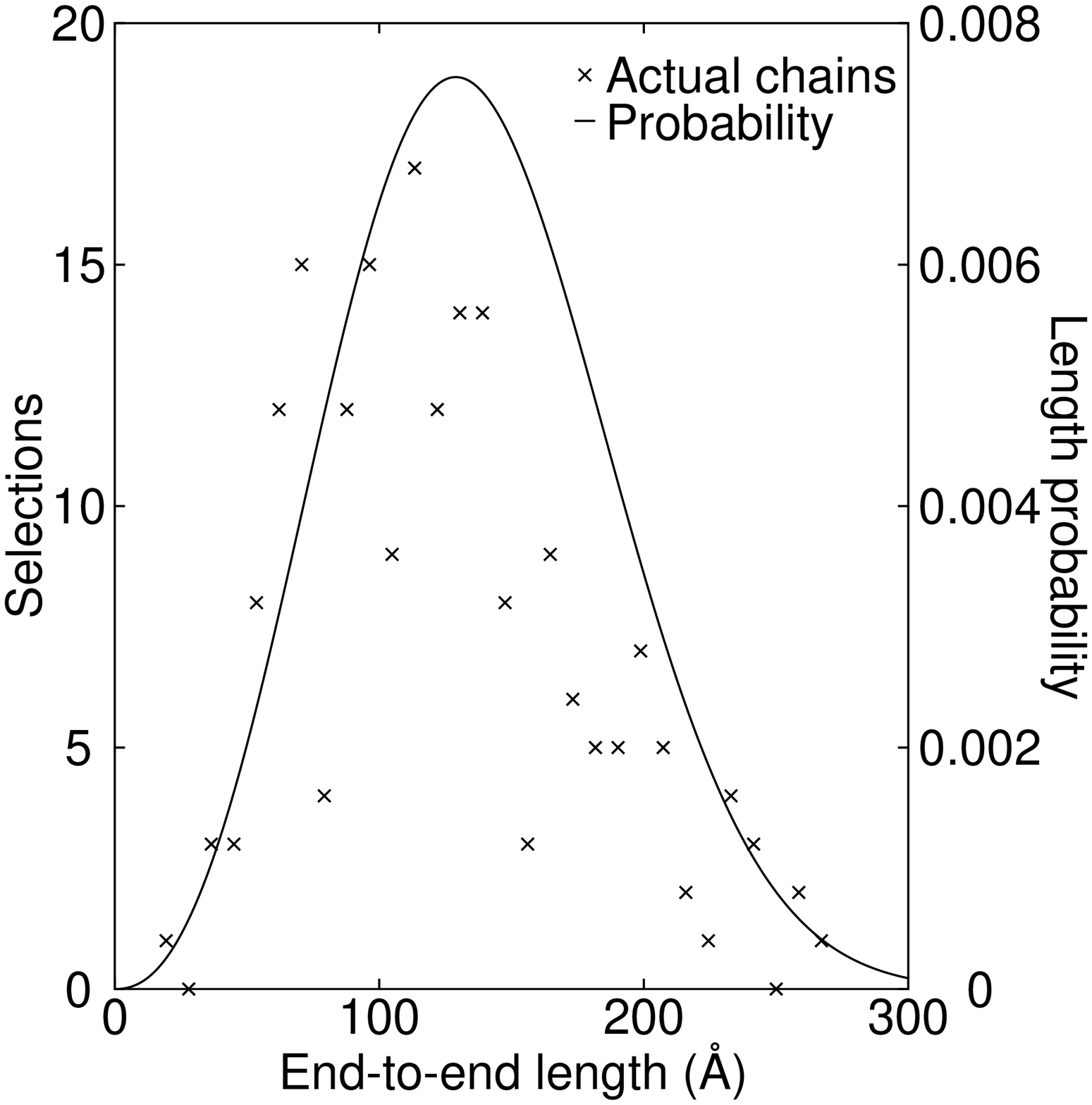}}
\caption{\label{fig_sa_length_histo}End-to-end length histogram for 200 chains 
of self-avoiding isotactic PMMA with 500 monomers each}
\end{center}
\end{figure}

The distribution of end-to-end chain lengths, shown in Figure \ref{fig_sa_length_histo}, follows the trend of the expected distribution \cite{ref_POLYMER_TEXT} for self-avoiding chains, 
% ---------------------------------------------------------
\begin{equation}
P(x)4\pi x^2dx = 4\pi x^2 \frac{0.278}{\sqrt{\left<R^2\right>}}x^{0.28}e^{-1.206x^{2.43}} dx,
\end{equation}
% ---------------------------------------------------------
which is the probability of finding the end of a self-avoiding chain of length 
$x = R/\sqrt{\left<R^2\right>}$ in volume $4\pi x^2dx$.
% ---------------------------------------------------------
% ---------------------------------------------------------

% -----------------------------------------------------------------------------
\section{Structure relaxation and molecular dynamics}
\label{sec_dynamics}

After the desired atomistic structure is built using the configurational biased direct Monte Carlo method it needs to be relaxed to remove close contacts. The tool performs a series of relaxations involving scaled-down van der Waals interactions that accomplish this task and also can ameliorate entanglement problems. Following these relaxations users can perform MD simulations to predict a variety of polymer properties including transport, glass transition temperature, and mechanical response.  \polymod\ allows users to relax the structures generated and perform MD simulations with the Dreiding potential and the LAMMPS simulation code. After the simulation is performed all the key data is displayed back in the user's web browser for analysis or download. This section describes the generation of an energy expression to compute total energies, stress, and forces that enable relaxation and dynamics followed by the methods to relax structures and options for MD simulations on the resulting atomistic models.

\subsection{Interatomic potential}
\label{sec_potential}
A general expression of the total potential energy in molecular force fields is obtained as a sum of energies due to valence or bonded interactions and non-bonded interactions:
\begin{equation}\label{EneEx}
U = \sum U_r + \sum U_{\theta} + \sum U_{\phi} + \sum U_{\omega} + \sum U_{vdw} + \sum U_e 
\end{equation}
where $U_r$ represents bond stretch interactions, $U_{\theta}$ bond angle bending, $U_{\phi}$ represents dihedral angle contributions, $U_{\omega}$ improper (out-of-plane) torsions, $U_{vdw}$ corresponds to nonbonded van der Waals interactions, and $U_e$ is electrostatic interactions originating from partial atomic charges. A variety of force fields are available for polymer simulations with specific parameterizations and functional forms. \polymod\  uses the DREIDING \cite{ref_DREIDING} force field, due to its wide applicability despite the relatively small number of parameters and because it is well validated for polymer chemistry. Partial atomic chargers are obtained with the Gasteiger approach \cite{ref_GASTEIGER1,ref_GASTEIGER2}. The following subsections describe the \polymod\ implementation of the energy expression to molecular mechanics and dynamics simulations. 

\subsection{Atom types}
Dreiding\cite{ref_DREIDING} and other molecular force fields use atom types to distinguish between like atoms in different bonding environments and setup the energy press ion in Eq. \ref{EneEx}. For example, the bond stretch, angle bending and torsional potentials are different for sp$^2$ and sp$^3$ C atoms. Dreiding uses four characters to specify atom types, the first two being the element name (adding ``\_''  for elements represented by a single character). The third character denotes hybridization or local bonding (2 means sp$^2$, 3 means sp$^3$, and R denotes resonance) and the fourth is used for atoms with implicit hydrogens. The atom type is determined by the element and the connectivity of a given atom and \polymod\ automatically types atoms in the monomers inputs by the users. If desired, users can specify Dreiding atom types when specifying monomers using the z-matrix format.

\subsubsection{Covalent interactions}
Covalent energies in Dreiding include bond stretch, angle bending, torsions involving dihedral angles, and improper torsions.  Identifying the terms corresponding to these interactions require bond connectivities. \polymod\  obtains connectivity information from the monomer z-matrix and checks for additional bonds with a distance-based criterion using atomic covalent radii. 
\par
The procedure for generating covalent interaction is briefly described as follows:
\begin{itemize}
\item The list of bond stretch terms is obtained from the builder and used to generate the energy expression terms as described below.
\item Bond angles are identified by listing bond pairs with a common atom central atom;
\item Dihedral angles $ijkl$ are obtained for each bond $jk$ by identifying atoms bonded to each atom in the central bond, based on bond information and atom linkage. 
\item Improper torsions are identified based on bond information and atom linkage. An improper term $ijkl$ is identified when three atoms $jkl$ are bonded to a central one $i$.
\end{itemize}
\par
Once all individual terms are identified \polymod\ assigns Dreiding force field parameters to each. Harmonic bonds are used with parameters depending on the two atoms making a bond.  Cosine-harmonic angle terms are used with parameters depending only on the type of central atom. The parameters describing torsion interactions in Dreiding depend on the hybridization of the central two atoms and the nature of these atoms is used to setup the energy expressions. Exceptions to this rule involve some torsions where the third atom is not an sp$^2$ center and in those cases the force field types of three atoms are used. The parameters for improper torsions are straightforward since they depend solely on the central atom of the improper term. 
\par
\subsubsection{Electrostatic interactions}
Partial atomic charges in the \polymod\ tool are obtained for all atoms via the iterative algorithm  of Gasteiger and Marsili \cite{ref_GASTEIGER1,ref_GASTEIGER2} that uses a partial equalization of orbital electronegativity.  These charges are calculated in a self-consistent manner from the ionization potentials and electron affinities of the neutral atoms and of their corresponding single charge cations, and they depend only on the connectivity of the atoms involved. A total charge of zero is maintained for each chain. A refinement  of the charge calculation was introduced in Ref. \cite{ref_GASTEIGER3} for structures with resonant $\pi$ bonds, such as aromatic rings. This is not yet  implemented in \polymod. 

Users can choose to describe long-range electrostatics via the particle-particle particle-mesh (pppm), the recommended method, or the Ewald method as implemented in LAMMPS. Alternative, a real-space cutoff can be used to describe electrostatics.

\subsection{van der Waals interactions}
Both Lennard-Jones and exponential-6 (Buckingham) functions can be used for van der Waals interactions in Dreding, and in \polymod\ MD simulations are performed using the exponential-6 functions since they provide a more accurate description \cite{ref_DREIDING,Li:2010p7210,Li:2011p7385}.  Lennard-Jones van der Waals functions are used, however, for the initial relaxation of the structure created by the builder since, unlike the exponential-6 function, it leads to repulsive interactions even for short interatomic distances.

\subsection{Structural relaxation}
A system created by the amorphous builder must be relaxed before MD calculations can be performed to remove close contacts that result in very large forces.  
Generalizing the method of Theodorou and Suter \cite{ref_THEODOROU}, the \polymod\ tool relaxes a system in multiple, successive conjugate gradient minimizations of the total energy. Scaled Dreiding Lennard-Jones parameters are used for this relaxation.  
Each minimization scales both the energy and distance parameters.  The use of scaled van der Waals interactions is motivated by the reduction in entanglements 
as chains can slide past one another as is done in state-of-the-art methods to relax amorphous polymers \cite{Li:2011p10031}. In addition, a gradual application of 
non-bond interactions has been shown to avoid unphysical changes in chain correlation functions \cite{ref_AUHL} as will be discussed in more detail
in Section \ref{sec_MD}.
\par
\begin{figure}[h]\begin{center}
\scalebox{0.5}[0.5]{\includegraphics{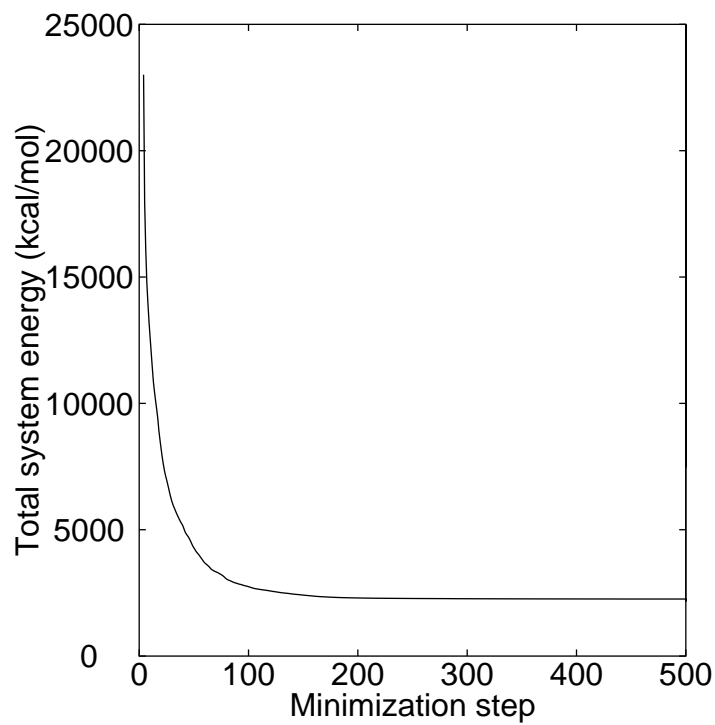}}
\caption{\label{fig_emin}Energy minimzation in LAMMPS of a single atactic 
PMMA chain with 96 monomers}
\end{center}
\end{figure}

Figure \ref{fig_emin} shows the energy of a first stage minimization, using LAMMPS, of an atactic PMMA chain of 96 monomers, built with the \polymod\ amorphous builder at a density of 0.7 g/cm$^3$.  Energies corresponding to steps 4 through 500 of the minimization are shown; the first 3 steps are omitted for scale, as the energies of those steps are significantly larger.
The relaxed system is then ready for MD, using unscaled Dreiding exponential-6 van der Waals interactions.

\subsection{Molecular dynamics}
\label{sec_MD}
After building and relaxing the structure is ready for MD simulations and property calculations.  \polymod\ provides users a large degree of control over the MD simulations that can be automatically performed within the tool using \nanohub.org computational resources; users can also use \polymod\ to generate LAMMPS input files and perform the simulations elsewhere. 
\par
The \polymod\ tool presents the user with a choice of MD ensembles, including 
NVT, NPT, and NVE.   The following options are also configurable:
\begin{itemize}
\item Time step
\item Number of steps
\item Temperature
\item Temperature change per time step
\item System pressure
\end{itemize}
\par
The user may also choose to deform the system by specifying the engineering strain rate, true strain rate, or applied stress along each Cartesian axis. This is useful to characterize the mechanical response of the structures created via non-equilibrium MD simulations.

Other MD options include the frequency of outputs, plotting the density as a function of time, and an initial thermalization prior to any MD steps. Furthermore, \polymod\ presents the option to run LAMMPS in serial mode on a single processor core or in parallel up to 64 cores on \nanohub\ HPC resources. The tool displays the progress of the MD simulation in the web browser window.
\par
After the simulation is performed the results are displayed graphically in the user's web browser.  These fully interactive outputs include:
\begin{itemize}
\item Structure produced by the amorphous builder, the input to the MD simulation
\item An animation of two monomers rotating about the backbone bond between them
\item LAMMPS input file (command script)
\item LAMMPS data file
\item Plot of system energy vs. conjugate gradient step during minimization prior to MD
\item Plot of total and potential energies vs. time during MD simulation
\item Plot of kinetic energy vs. time during MD simulation
\item Plot of temperature vs. time during MD simulation
\item Diagonal components of the stress tensor vs. time during MD simulation
\item Off-diagonal (shear) components of the stress tensor vs. time during MD simulation
\item Animation of atoms during relaxation and MD simulation
\item Output log containing all text output from the amorphous builder and LAMMPS
\end{itemize}
All outputs can be downloaded as an image or as comma-separated text, in the case of plots.
\par
\begin{figure}[h]\begin{center}
\scalebox{0.5}[0.5]{\includegraphics{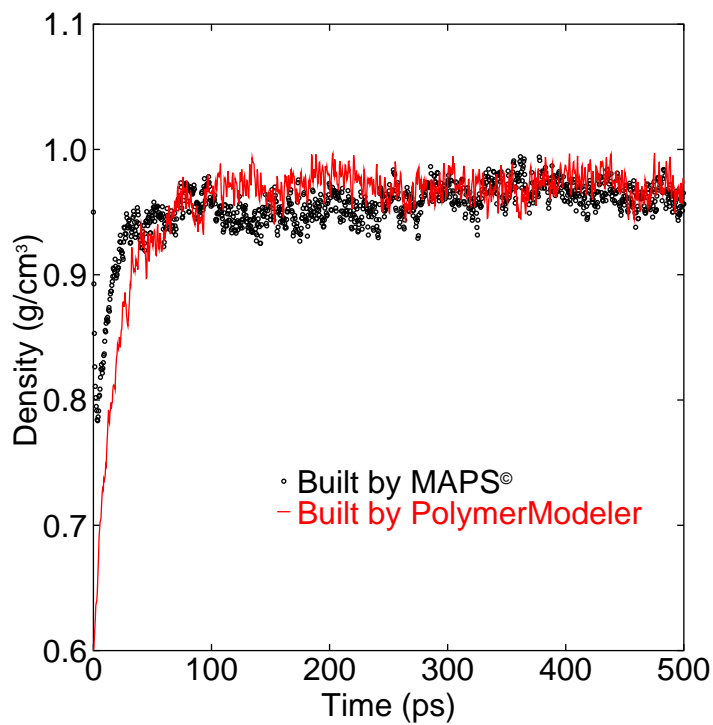}}
\caption{\label{fig_density}Time evolution of density for 5 PMMA chains (1 atactic, 4 syndiotactic)}
\end{center}
\end{figure}
Figure \ref{fig_density} shows a comparison of two systems, each with 5 PMMA chains with 96 monomers in each chain; 1 chain was atactic and 4 were syndiotactic.  One system was built at an initially high density with the commercial package MAPS, and the other was built at an initially low density with the \polymod\ amorphous builder.  The built systems were then relaxed using LAMMPS (separate from the \polymod\ tool) in NPT simulations at 600K for 500 ps with a time step of 1 fs.  The trend in density over time is very similar in both cases, indicating that the system built by \polymod\ is as realistic as the system built by MAPS.

\section{Using \polymod\ to predict the structure and properties of amorphous polymers}

\subsection{Molecular structure of polyethylene}

The molecular relaxation timescales of polymer melts (which increase with molecular weight as $N^{3.4}$ for 
high molecular weight cases\cite{RubinsteinColby}) are prohibitively long for molecular dynamics simulations
in most cases of interest. Thus the initial molecular structure and its relaxation is critical for accurate MD simulations. 
\cite{ref_HOY,ref_AUHL} The mean square displacement (MSD) between monomers as a function of their index in 
a polymer chain was found to be a useful descriptor of the polymer structure. Auhl et al. \cite{ref_AUHL} showed that a simple
relaxation of as-built polymer samples via energy minimization leads to unphysical molecular structure 
changes that relax over very long timescales and, consequently, affect predictions. These changes are 
reflected as significant deviations in the MSD for short and intermediate distances. They found that a {\it slow push-off} 
approach, where intermolecular forces are turned on gradually, minimizes these problems and leads to well-equilibrated
molecular structures.

\begin{figure}[h]\begin{center}
\scalebox{0.5}[0.5]{\includegraphics{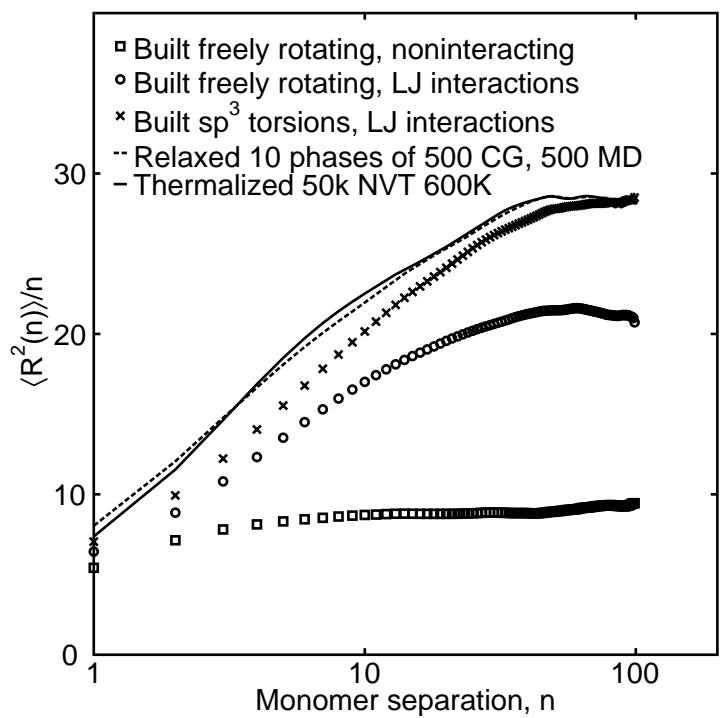}}
\caption{\label{fig_msid}Mean square internal distances between 
monomers for 100 chains of polyethylene built at 1.0 g/cm$^3$.  Built systems
were then relaxed and thermalized.}
\end{center}
\end{figure}

\polymod\ allows multi-step relaxations where non-bond interactions are turned on gradually and we
subjected our model to the same tests as in Ref. \cite{ref_AUHL}. Figure \ref{fig_msid} shows the mean square internal 
distances (MSD) between monomers for chains of polyethylene built using various algorithms, for comparison with 
Ref. \cite{ref_AUHL}. As expected, building the chains using a naive, freely-rotating model without avoiding close 
contacts leads to the smallest mean square displacements; adding Lennard Jones interactions
during build increases the MSD and using the configuration biased Monte Carlo approach including
van der Waals interactions and the torsional barriers caused by covalent terms leads to an even higher
MSD and a more accurate structure.  For this last cases, the system was built with sp$^3$ torsional potential
from Dreiding (Equation 13 in reference \cite{ref_DREIDING}) and Lennard-Jones interactions between 
all atoms to a cutoff of 6{\AA}. The as-built structure was then relaxed in a gradual, 10-step, procedure where van
der Waals interactions are turned on slowly. At each step we performed 500 conjugate gradient steps and 500 
molecular dynamics steps at a temperature T=600 K. The properties of the resulting structure are shown as dashed lines in 
Fig. \ref{fig_msid}. Finally, we performed an additional thermalization with MD for 50,000 steps under isothermal, isochoric 
conditions at 600 K.  The MSD of the structure after the MD simulations is shown as a solid line in Fig. \ref{fig_msid}. 
Consistent with the findings by Auhl et al. we find that a gradual relaxation procedure results in structures that 
maintain their statistical properties during relaxation and dynamics. Such structures produced by the \polymod\ builder 
with sp$^3$ torsions and Lennard-Jones interactions followed by a gradual introduction of non-bond interactions are 
expected to be good starting points for MD simulations.  

\begin{figure}[h]\begin{center}
\scalebox{0.5}[0.5]{\includegraphics{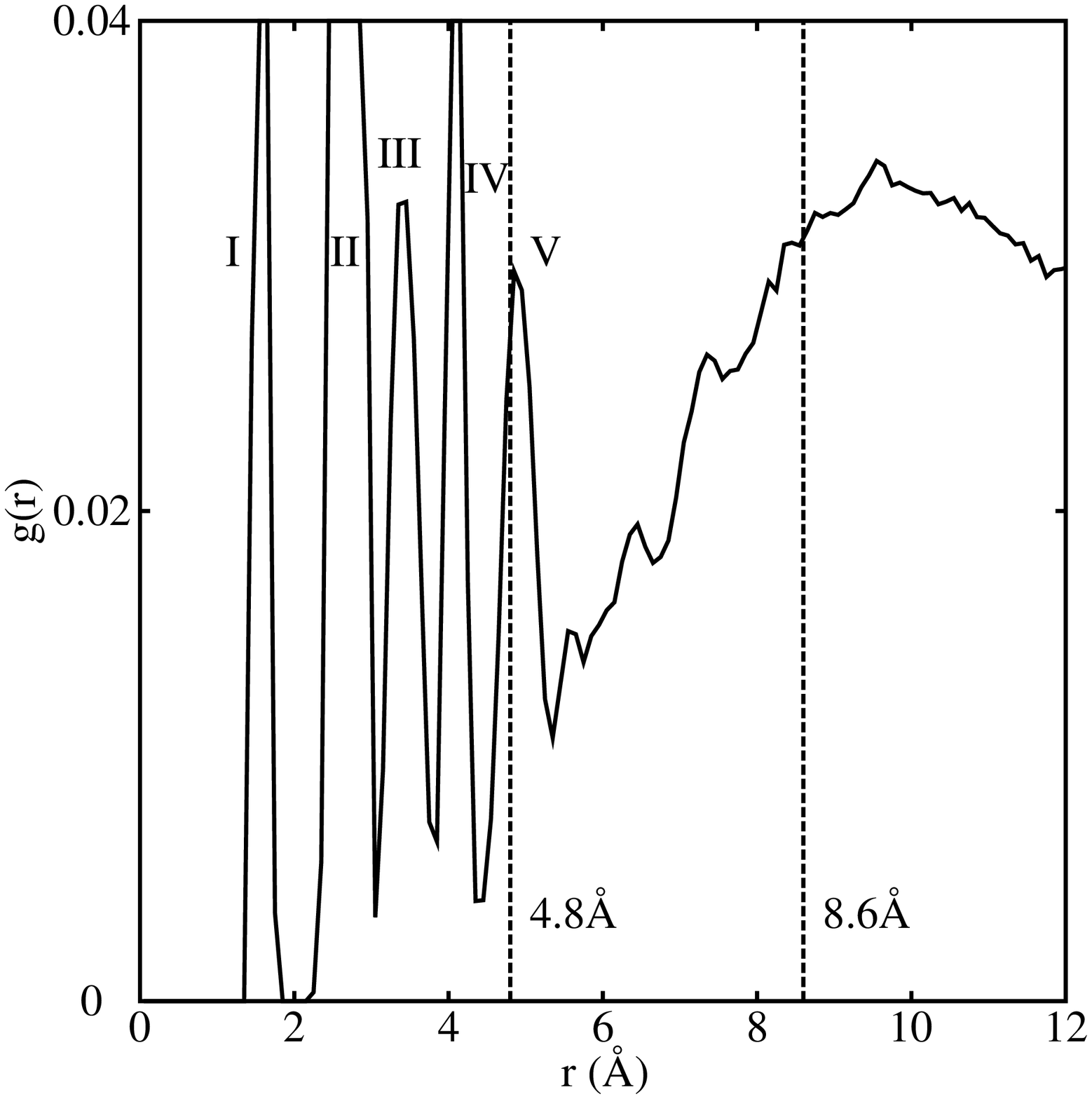}}
\caption{\label{fig_gr}Radial distribution function for PMMA backbone carbon
atoms for 100 chains of isotactic PMMA.}
\end{center}
\end{figure}

In addition to the MSD, the radial distribution function g(r) for chains constructed by \polymod\ demonstrates good
equilibration and good agreement with neutron diffraction experiments.  Figure \ref{fig_gr} shows the radial distribution function, 
g(r), for backbone carbon atoms in 100 isotactic PMMA chains built at 0.6 g/cm$^3$, fully relaxed with the 10 steps of minimization, 
after 400 ps of MD at 600K.  The peaks corresponding to nearest neighbor distances are labeled.  Indicators at 4.8{\AA} and 8.6{\AA} are 
included in figure \ref{fig_gr} for comparison with Figure 7(a) in Ref. \cite{Genix:2006p3947}, which shows a similar g(r) calculation for 
the backbone atoms of syndiotactic PMMA.  The indicators correspond to intrachain peaks (4.8{\AA}) and interchain peaks (8.6{\AA}) 
identified in PMMA by neutron scattering data. Structures built  and equilibrated by \polymod\ compare well to these results.

We encourage \polymod\ users to perform similar tests of their structures before production runs. We also note that
the structures generated can used as starting points for advanced parallel-replica type approaches specifically designed
to equilibrate long-chain amorphous structures \cite{Li:2011p10031}.

% ---------------------------------------------------------
% ---------------------------------------------------------

% ---------------------------------------------------------
% ---------------------------------------------------------

\subsection{Prediction of the glass transition temperature of polymthylmethacrylate}

In this section we illustrate the prediction of the glass transition temperature (Tg) of polymthylmethacrylate (PMMA) using MD over a structure built with \polymod\. We created a molecular structure using the continuous configurational biased direct Monte Carlo algorithm using torsional potentials corresponding to sp$^3$ bonds and a cutoff of 6 \AA\ to compute van der Waals interactions. The simulation cell contains 40 chains, each 90-monomers long, for a total of 54,080 atoms. The structure is then relaxed using a 3 phase minimization and thermalized for 100 ps under NPT conditions at 600 K.  In order to compute Tg the relaxed structure is cooled down to room temperature under constant pressure conditions. We control the temperature and pressure of the system using Nos\'{e}-Hoover thermostat and barostat with coupling timescales of 0.1 ps and 1 ps respectively. The system is cooled down in continuous increments at a rate of 0.1 K/ps. These simulations take approximately 70 hours when run using 24 cores.
% ---------------------------------------------------------
\begin{figure}[h]\begin{center}
\scalebox{0.5}[0.5]{\includegraphics{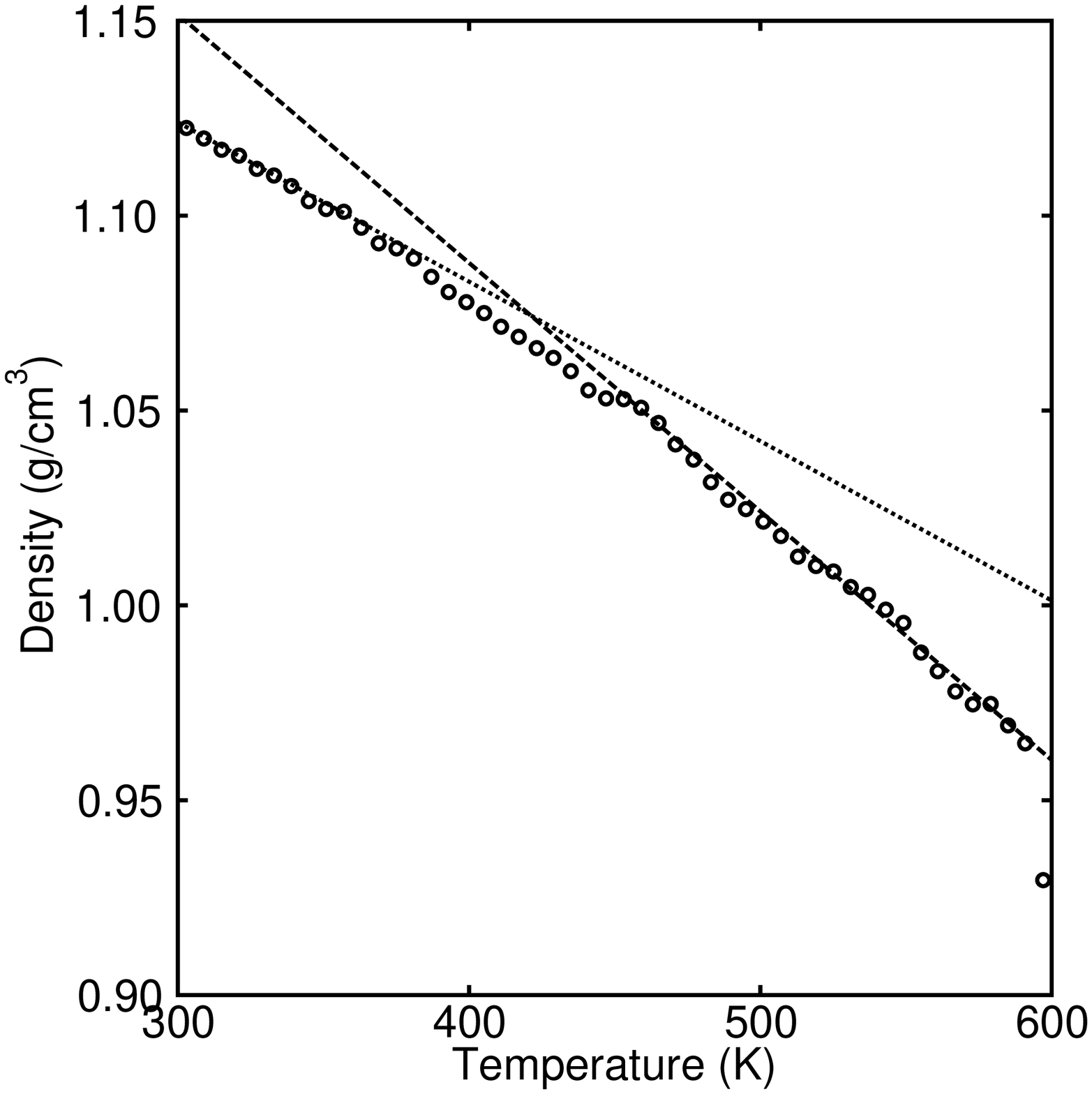}}
\caption{\label{fig_Tg}Glass transition while cooling 54,080 PMMA atoms.}
\end{center}
\end{figure}
% ---------------------------------------------------------

Figure \ref{fig_Tg} shows the resulting density as a function of temperature during cooling. The thermal contraction that results from the anharmonicity in the interaction potentials is clearly seen. More importantly for our study is the change is slope that can be observed around 420 K. This denotes the glass transition where the system transforms from a melt to a glassy solid. 
The molten state above Tg has a higher thermal expansion coefficient that the solid below it. The predicted value of Tg is in good agreement with experimental values 388 K\cite{ref_FOX} and 385 K\cite{ref_BURGUETE}. 
When comparing MD and experimental glass transition temperatures the effect of cooling/heating rate should be considered. The glass transition temperature increases with increasing heating/cooling rate and this increase is approximately 3K per order of magnitude increase in rate\cite{ref_FERRY,ref_LI_POLYMER53,ref_LI_POLYMER52}.

% -----------------------------------------------------------------------------
\section{Conclusions}
\label{sec_conclusions}
The \polymod\ tool at \nanohub.org features an amorphous builder that allows customized specification of monomers, including multiple monomers and monomer patterns to support stereochemistry.  Chain conformations are determined by specifying the torsion angle probabilities of backbone atoms.  The well-known cases of freely rotating and self-avoiding chains are accurately reproduced by the builder.  Structural properties of systems built by the \polymod\ builder compare well with systems built by commercial packages and current state-of-the-art techniques.
\par
After a system of polymer chains is built it may be relaxed, thermalized, deformed, and otherwise simulated using LAMMPS directly from the \polymod\ tool.  Large LAMMPS simulations run on HPC resources affiliated with \nanohub.  All simulations are launched directly from a standard web browser, and all results are displayed in the browser.  No additional software, installation, or remote accounts are necessary to use this free tool. The structures produced by  \polymod\ can also be used as starting points for additional relaxations using special-purpose algorithms \cite{Li:2011p10031}.
\par
Users can submit support tickets and ask questions in \nanohub\ forums.  The \polymod\ tool, like all \nanohub\ tools, features a wish list of requested features.  Future development is guided in part by requests from users via the wish list. Some of the features we would like to incorporate in \polymod\ are:
\begin{itemize}
\item Partial charge calculations using electronegativity equalization methods.
\item Implementing force fields other than Dreiding for MD simulations.
\item Gasteiger charges for resonant $\pi$ bonds
\end{itemize}
\par
We believe the combination of a powerful and flexible amorphous builder, the ability to perform LAMMPS-based MD simulations, the convenience of running simulations interactively through a web browser, and the user feedback options make \polymod\ a very compelling, free alternative to commercial polymer simulation packages. Combining power and ease of use, we foresee \polymod\ will be used both for research and education.

\section*{Acknowledgements}
This work was supported by the US National Science Foundation, CMMI Grant 
0826356 and by Purdue University.   The authors gratefully acknowledge 
computational resources from nanoHUB.org and support from the Boeing Co.
The authors also thank David Lowing for testing the glass transition calculations.

% -----------------------------------------------------------------------------
\bibliographystyle{unsrt}
%\bibliographystyle{elsarticle-num}
%\bibliography{refs}

% -----------------------------------------------------------------------------
\end{document}